\documentclass[fleqn,twoside]{article}
\usepackage{espcrc2}
\usepackage{graphicx}

\title{Supernovae and Neutrinos}
\author{John F. Beacom
\address{NASA/Fermilab Astrophysics Center, Fermilab,
Batavia, Illinois 60510-0500, USA \\ {\tt beacom@fnal.gov}}}

\begin{document}

\begin{abstract}
A long-standing problem in supernova physics is how to measure the
total energy and temperature of $\nu_\mu$, $\nu_\tau$,
$\bar{\nu}_\mu$, and $\bar{\nu}_\tau$.  While of the highest
importance, this is very difficult because these flavors only have
neutral-current detector interactions.  We propose that
neutrino-proton elastic scattering, $\nu + p \rightarrow \nu + p$, can
be used for the detection of supernova neutrinos in scintillator
detectors.  It should be emphasized immediately that the dominant
signal is on {\it free} protons.  Though the proton recoil kinetic
energy spectrum is soft, with $T_p \simeq 2 E_\nu^2/M_p$, and the
scintillation light output from slow, heavily ionizing protons is
quenched, the yield above a realistic threshold is nearly as large as
that from $\bar{\nu}_e + p \rightarrow e^+ + n$.  In addition, the
measured proton spectrum is related to the incident neutrino spectrum.
The ability to detect this signal would give detectors like KamLAND
and Borexino a crucial and unique role in the quest to detect
supernova neutrinos.  These results are now published: J.~F.~Beacom,
W.~M.~Farr and P.~Vogel, Phys.\ Rev.\ D {\bf 66}, 033001 (2002)
[arXiv:hep-ph/0205220]; the details are given there~\cite{elastic}.
\end{abstract}
\maketitle

%%%%%%%%%%%%%%%%%%%%%%%%%%%%%%%%%%%%%%%%%%%%%%%%%%%%%%%%%%%%%%%%%%%%%%%%%%%%%
%			Introduction					    %
%%%%%%%%%%%%%%%%%%%%%%%%%%%%%%%%%%%%%%%%%%%%%%%%%%%%%%%%%%%%%%%%%%%%%%%%%%%%%

\section{Introduction}

When the next Galactic supernova occurs, approximately $10^4$ detected
neutrino events are expected among the several detectors around the
world.  It is widely believed that these $10^4$ events will provide
important clues to the astrophysics of the supernova as well as the
properties of the neutrinos themselves.  Interestingly, recent
breakthroughs in understanding solar and atmospheric neutrinos each
occurred when the accumulated samples of detected events first exceeded
$10^4$.

But will we have enough information to study the supernova neutrino
signal in detail?  Almost all of the detected events will be
charged-current $\bar{\nu}_e + p \rightarrow e^+ + n$, which will be
well-measured, both because of the large yield and because the
measured positron spectrum is closely related to the neutrino
spectrum. Because of the charged-lepton thresholds, the flavors
$\nu_\mu$, $\nu_\tau$, $\bar{\nu}_\mu$, and $\bar{\nu}_\tau$ can only
be detected in neutral-current reactions, of which the total yield is
expected to be approximately $10^3$ events.  However, in general one
{\it cannot} measure the neutrino energy in neutral-current reactions.
This talk presents an exception.  These four flavors are expected to
carry away about 2/3 of the supernova binding energy, and are expected
to have a higher temperature than $\nu_e$ or $\bar{\nu}_e$.  However,
there is no experimental basis for these statements, and present
numerical models of supernovae cannot definitively address these
issues either.  If there is no spectral signature for the
neutral-current detection reactions, then neither the total energy
carried by these flavors nor their temperature can be separately
determined from the detected number of events.

But it is crucial that these quantities be {\it measured}.  Both are
needed for comparison to numerical supernova models.  The total energy
is needed to determine the mass of the neutron star, and the
temperature is needed for studies of neutrino oscillations.  At
present, such studies would suffer from the need to make
model-dependent assumptions.  This problem has long been known, but
perhaps not widely enough appreciated.  In this talk, I clarify this
problem, and provide a realistic solution that can be implemented in
two detectors, KamLAND (already operating) and Borexino (to be
operating soon).  The solution is based on neutrino-proton elastic
scattering, which has never before been shown to be a realistic
detection channel for low-energy neutrinos.

In this talk, I will focus on just the problem of measuring the
temperature and total energy of $\nu_\mu$, $\nu_\tau$,
$\bar{\nu}_\mu$, and $\bar{\nu}_\tau$, since everything else in
understanding supernova neutrinos depends on it.

%%%%%%%%%%%%%%%%%%%%%%%%%%%%%%%%%%%%%%%%%%%%%%%%%%%%%%%%%%%%%%%%%%%%%%%%%%%%
%			Cross Section					   %
%%%%%%%%%%%%%%%%%%%%%%%%%%%%%%%%%%%%%%%%%%%%%%%%%%%%%%%%%%%%%%%%%%%%%%%%%%%%

\section{Cross Section}
\label{cross}

The cross section for neutrino-proton elastic scattering is an
important prediction~\cite{weinberg} of the Standard Model, and it has
been confirmed by extensive measurements at GeV energies (see, e.g.,
Ref.~\cite{ahrens}).  At the energies considered here, the full cross
section formula~\cite{weinberg,ahrens,sigma} can be greatly
simplified.  The differential cross section as a function of neutrino
energy $E_\nu$ and struck proton recoil kinetic energy $T_p$ (and mass
$M_p$) is
\begin{eqnarray}
\frac{d\sigma}{dT_p} & = & \frac{G_F^2 M_p}{\pi} \\
& \times & \!\!\!\! 
\left[\left(1 - \frac{M_p T_p}{2 E_\nu^2}\right) c_V^2 +
\left(1 + \frac{M_p T_p}{2 E_\nu^2}\right) c_A^2\right]\,. \nonumber
\label{diffxs}
\end{eqnarray}
In this equation, we have taken $(E_\nu - T_p)^2 \simeq E_\nu^2$
(i.e., keeping only the lowest order in $E_\nu/M_p$, a very good
approximation); the full expression was used in the calculations
below.  The neutral-current coupling constants between the exchanged
$Z^\circ$ and the proton are
\begin{eqnarray}
c_V & = & \frac{1 - 4\sin^2\theta_w}{2} = 0.04\,, \\ 
c_A & = & \frac{1.27}{2}\,,
\end{eqnarray}
where the factor 1.27 is determined by neutron beta decay.  The cross
section for antineutrinos is obtained by the substitution $c_A
\rightarrow -c_A$.  As will be emphasized below, our results are
totally independent of oscillations among active flavors, as this is a
neutral-current reaction.

We consider only {\it free} proton targets; the small yield from bound
protons creates a small background signal, as discussed in
Ref.~\cite{elastic}.  We use the struck proton kinetic energy in the
laboratory frame as our kinematic variable, as appropriate to
scintillator detectors.  For a neutrino energy $E_\nu$, $T_p$ ranges
between 0 and $T_p^{max}$, where
\begin{equation}
T_p^{max} = \frac{2 E_\nu^2}{M_p + 2 E_\nu} 
\simeq \frac{2 E_\nu^2}{M_p} \,.
\end{equation}
The maximum is obtained when the neutrino recoils backwards with its
original momentum $E_\nu$, and thus the proton goes forward with
momentum $2 E_\nu$.  Since $c_A \gg c_V$, the {\it largest} proton
recoils are favored, which is optimal for detection.

%%%%%%%%%%%%%%%%%%%%%%%%%%%%%%%%%%%%%%%%%%%%%%%%%%%%%%%%%%%%%%%%%%%%%%%%%%%%%
%			Supernova Neutrinos				    %
%%%%%%%%%%%%%%%%%%%%%%%%%%%%%%%%%%%%%%%%%%%%%%%%%%%%%%%%%%%%%%%%%%%%%%%%%%%%%

\section{Supernova Neutrinos}

In this talk, I characterize the supernova neutrino signal in a very
simple way, though consistently with numerical supernova
models~\cite{SNmodels}.  The change in gravitational binding energy
between the initial stellar core and the final proto-neutron star is
about $3 \times 10^{53}$ ergs, about $99\%$ of which is carried off by
all flavors of neutrinos and antineutrinos over about 10 s.  The
emission time is much longer than the light-crossing time of the
proto-neutron star because the neutrinos are trapped and must diffuse
out, eventually escaping with approximately Fermi-Dirac spectra
characteristic of the surface of last scattering.  In the usual model,
$\nu_\mu$, $\nu_\tau$ and their antiparticles are emitted with
temperature $T \simeq 8$ MeV, $\bar{\nu}_e$ has $T \simeq 5$ MeV, and
$\nu_e$ has $T \simeq 3.5$ MeV.  The temperatures differ from each
other because $\bar{\nu}_e$ and $\nu_e$ have charged-current opacities
(in addition to the neutral-current opacities common to all flavors),
and because the proto-neutron star has more neutrons than protons.  It
is generally assumed that each of the six types of neutrino and
antineutrino carries away about $1/6$ of the total binding energy,
though this has an uncertainty of at least $50\%$~\cite{raffeltproc}.
The supernova rate in our Galaxy is estimated to be $(3 \pm 1)$ per
century (this is reviewed in Ref.~\cite{SNrate}).

The expected number of events (assuming a hydrogen to carbon ratio of
$2:1$) is 
\begin{eqnarray}
\label{yield}
N & = &
70.8
\left[\frac{E}{10^{53}{\rm\ erg}}\right]
\left[\frac{1{\rm\ MeV}}{T}\right] \nonumber \\ 
& \times &
\left[\frac{10{\rm\ kpc}}{D}\right]^2
\left[\frac{M_D}{1{\rm\ kton}}\right]
\left[\frac{\langle \sigma \rangle}{10^{-42}{\rm\ cm^2}}\right]\,.
\end{eqnarray}
(Though written slightly differently, this is equivalent to
the similar expression in Ref.~\cite{SNmb}.)
We assume $D = 10$ kpc, and a detector fiducial mass of 1 kton for KamLAND.
As written, Eq.~(\ref{yield}) is for the yield per flavor, assuming
that each carries away a portion $E$ of the total binding energy
(nominally, $E_B = 3 \times 10^{53}$ ergs, and $E = E_B/6$).
The thermally-averaged cross section (the integral of the cross
section with normalized Fermi-Dirac distribution) is defined for each
${\rm C H}_2$ ``molecule'', and a factor of 2 must be included for
electron or free proton targets.  

Prior to Ref.~\cite{elastic}, the largest expected yield in any oil or
water detector was from $\bar{\nu}_e + p \rightarrow e^+ + n$.  The
total cross sections for charged-current $\bar{\nu}_e + p \rightarrow
e^+ + n$ and neutral-current $\nu + p \rightarrow \nu + p$ have
similar forms, though the latter is about 4 times smaller.  However,
this is compensated in the yield by the contributions of all six
flavors, as well as the higher temperature assumed for $\nu_\mu$ and
$\nu_\tau$ ($T = 8$ MeV instead of 5 MeV).  Thus, the total yield from
$\nu + p \rightarrow \nu + p$ is {\it larger} than that from
$\bar{\nu}_e + p \rightarrow e^+ + n$, when the detector threshold is
neglected.  

Taking into account radiative, recoil, and weak magnetism
corrections, the thermally-averaged cross section for $\bar{\nu}_e + p
\rightarrow e^+ + n$ at $T = 5$ MeV is $44 \times 10^{-42}$ cm$^2$
(for 2 protons)~\cite{invbeta}.  These corrections reduce the
thermally-averaged cross section by about 20\%, and also correct the
relation $E_e = E_\nu - 1.3$ MeV.  The total expected yield from this
reaction is thus about 310 events in 1 kton.

Since the struck protons in $\nu + p \rightarrow \nu + p$ have a
relatively low-energy recoil spectrum, and since realistic detectors
have thresholds, it is crucial to consider the proton spectrum in
detail, and not just the total yield of neutrinos that interact.

%%%%%%%%%%%%%%%%%%%%%%%%%%%%%%%%%%%%%%%%%%%%%%%%%%%%%%%%%%%%%%%%%%%%%%%%%%%%%
%			Proton Spectrum					    %
%%%%%%%%%%%%%%%%%%%%%%%%%%%%%%%%%%%%%%%%%%%%%%%%%%%%%%%%%%%%%%%%%%%%%%%%%%%%%

\section{Proton Recoil Spectrum}

The elastically-scattered protons will have kinetic energies of a few
MeV.  Obviously, these nonrelativistic protons will be completely
invisible in any \v{C}erenkov detector like Super-Kamiokande.
However, such small energy depositions can be readily detected in
scintillator detectors such as KamLAND and Borexino.  We first
consider the true proton spectrum, and then how this spectrum would
appear in a realistic detector.

%%%%%%%%%%%%%%%%%%%%
\begin{figure}
\centerline{\includegraphics[width=18pc]{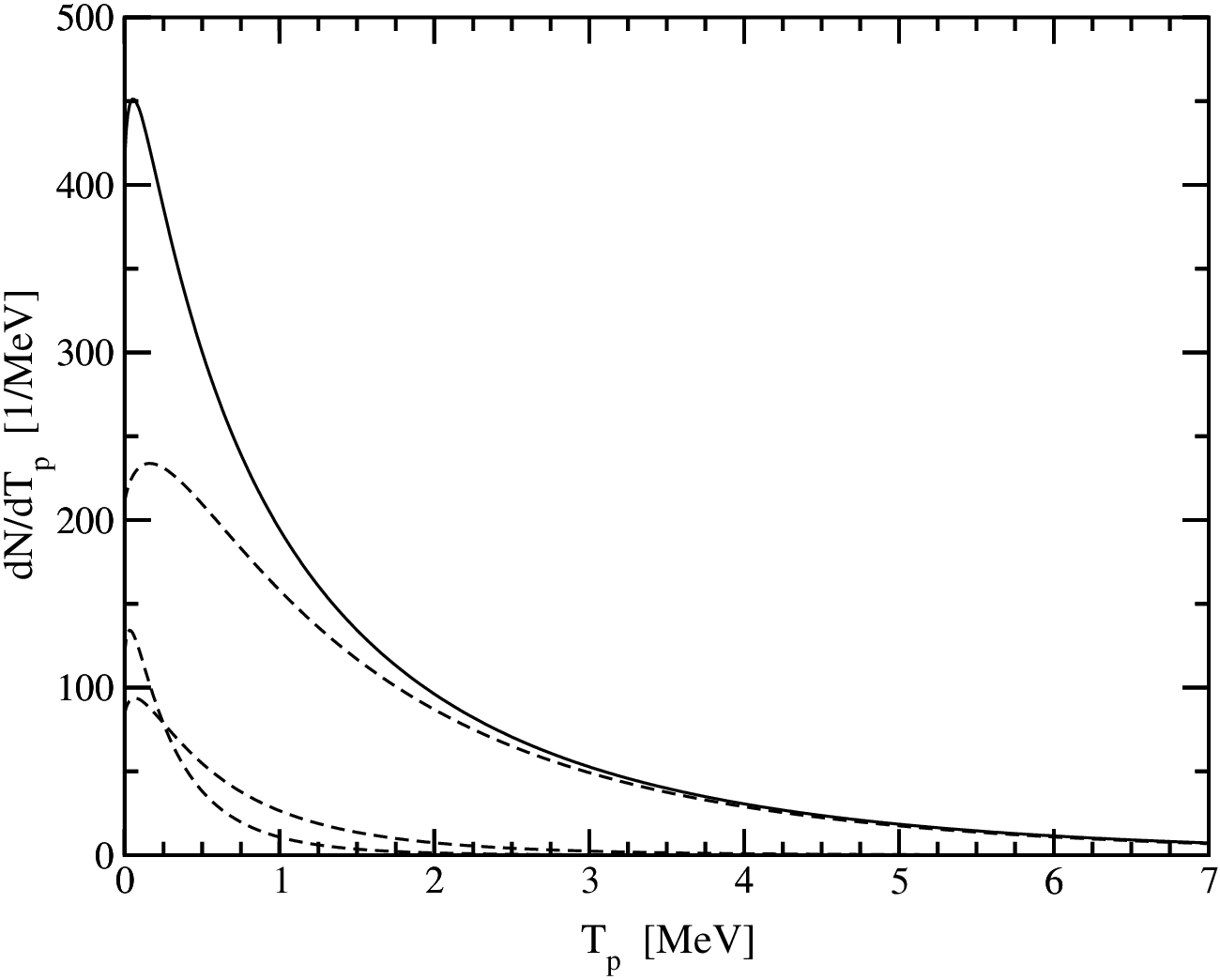}}\vspace{-1cm}
\caption{\label{fig:dndt} The true proton spectrum in KamLAND, for a
standard supernova at 10 kpc.  In order of increasing maximum kinetic
energy, the contributions from $\nu_e$, $\bar{\nu}_e$, and the sum of
$\nu_\mu$, $\nu_{\tau}$, $\bar{\nu}_{\mu}$, and $\bar{\nu}_{\tau}$ are
shown with dashed lines.  The solid line is the sum spectrum for all
flavors.  Taking the detector properties into account substantially
modifies these results, as shown below.}
\end{figure}
%%%%%%%%%%%%%%%%%%%%

The proton spectrum (for one flavor of neutrino) is given by
\begin{equation}
\label{pspec}
\frac{dN}{dT_p}\left(T_p\right) =
C \int_{(E_\nu)_{min}}^{\infty} \!\!\!\!
{dE_\nu\, f(E_\nu) \, \frac{d\sigma}{dT_p}\left(E_\nu, T_p\right)}\,,
\end{equation}
where $f(E_\nu)$ is a normalized Fermi-Dirac spectrum and the
differential cross section is given by Eq.~(\ref{diffxs}).  For a
given $T_p$, the minimum required neutrino energy is
\begin{equation}
(E_\nu)_{min} = \frac{T_p + \sqrt{T_p (T_p + 2 M_p)}}{2} 
\simeq \sqrt{\frac{M_p T_p}{2}}\,.
\end{equation}
The normalization constant $C$ is determined by Eq.~(\ref{yield}),
as the integral of Eq.~(\ref{pspec}) over all $T_p$ without the $C$ 
factor is $\langle \sigma \rangle$.

Throughout, I refer to the $\nu_e$ ($T = 3.5$ MeV), $\bar{\nu}_e$ ($T
= 5$ MeV), and the combined $\nu_\mu$, $\nu_{\tau}$,
$\bar{\nu}_{\mu}$, and $\bar{\nu}_{\tau}$ ($T = 8$ MeV) flavors.
Since we know that there are neutrino oscillations, this language is
somewhat incorrect.  However, our results are {\it totally
insensitive} to any oscillations among active neutrinos or
antineutrinos (since this is a neutral-current cross section), and
also to oscillations between active neutrinos and antineutrinos (since
the cross section is dominated by the $c_A^2$ terms).  Thus when we
refer to the $\nu_e$ flavor, we mean ``those neutrinos emitted with a
temperature $T = 3.5$ MeV, whatever their flavor composition now,''
etc.  The true proton spectra corresponding to the various flavors are
shown in Fig.~\ref{fig:dndt}.  As seen in the figure, the
contributions of $\nu_e$ and $\bar{\nu}_e$ are quite suppressed
relative to the sum of $\nu_\mu$, $\nu_{\tau}$, $\bar{\nu}_{\mu}$, and
$\bar{\nu}_{\tau}$.

%%%%%%%%%%%%%%%%%%%%%%%%%%%%%%%%%%%%%%%%%%%%%%%%%%%%%%%%%%%%%%%%%%%%%%%%%%%%%
%				Quenching				    %
%%%%%%%%%%%%%%%%%%%%%%%%%%%%%%%%%%%%%%%%%%%%%%%%%%%%%%%%%%%%%%%%%%%%%%%%%%%%%

\section{Quenching}
\label{quenching}

\begin{figure}
\centerline{\includegraphics[width=18pc]{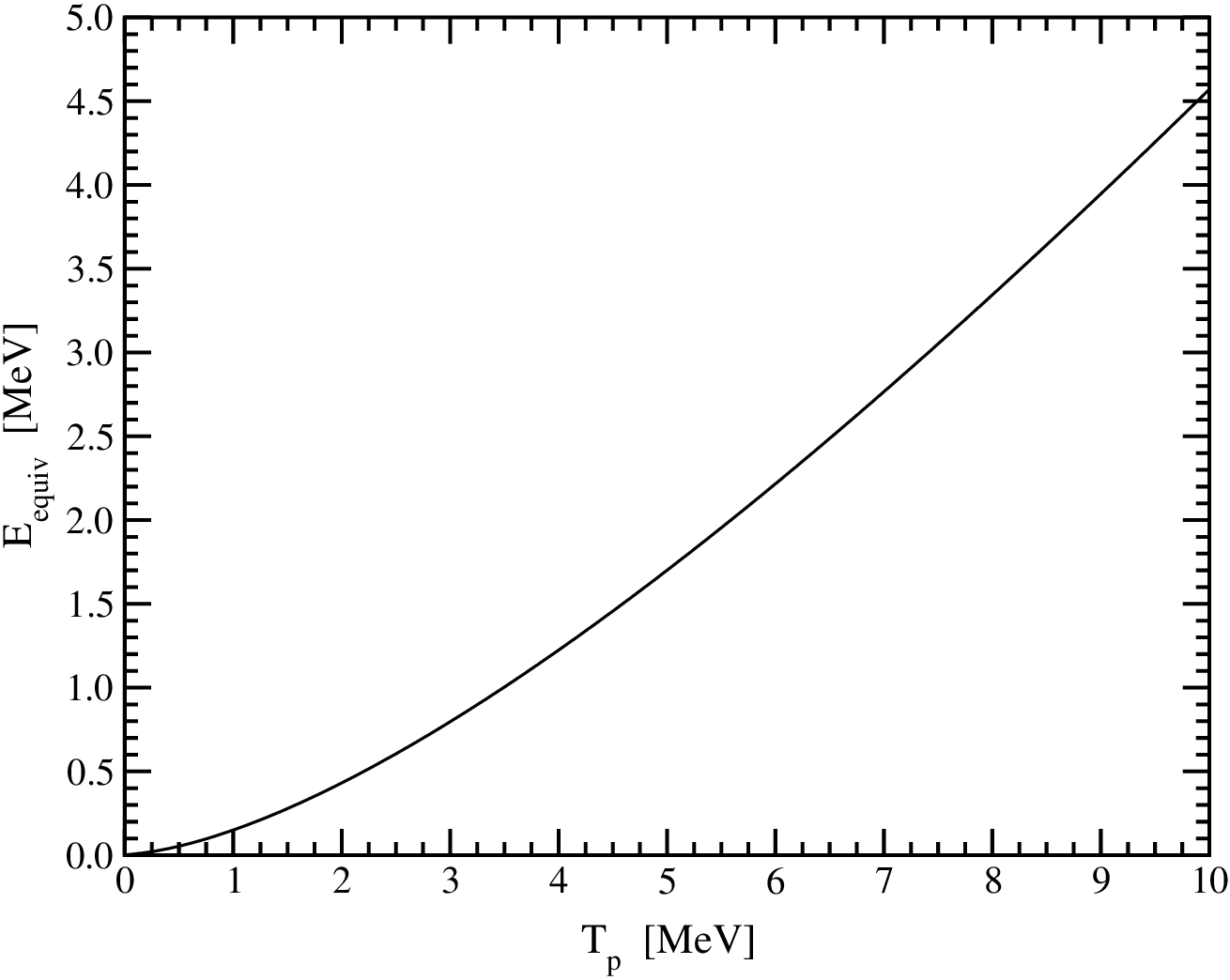}}\vspace{-1cm}
\caption{\label{fig:quench} The quenched energy deposit (equivalent
electron energy) as a function of the proton kinetic energy.  The
KamLAND detector properties are assumed.}
\end{figure}

For highly ionizing particles like low-energy protons, the light
output is reduced or ``quenched'' relative to the light output for an
electron depositing the same amount of energy.  The observable light
output $E_{equiv}$ (i.e., equivalent to an electron of energy
$E_{equiv}$) can be calculated by integrating Birk's Law with tables
of $dE/dx$ for protons in the KamLAND oil-scintillator
mixture~\cite{kamland}.  The observed energy in terms of the proton
kinetic energy is shown in Fig.~\ref{fig:quench}.  Thus the proton
quenching factor ($E_{equiv}/T_p$) is thus roughly 1/2 at 10 MeV, 1/3
at 6 MeV, 1/4 at 3 MeV, and so on.

%%%%%%%%%%%%%%%%%%%%
%%%%%%%%%%%%%%%%%%%%

%%%%%%%%%%%%%%%%%%%%%%
\begin{figure}
\centerline{\includegraphics[width=18pc]{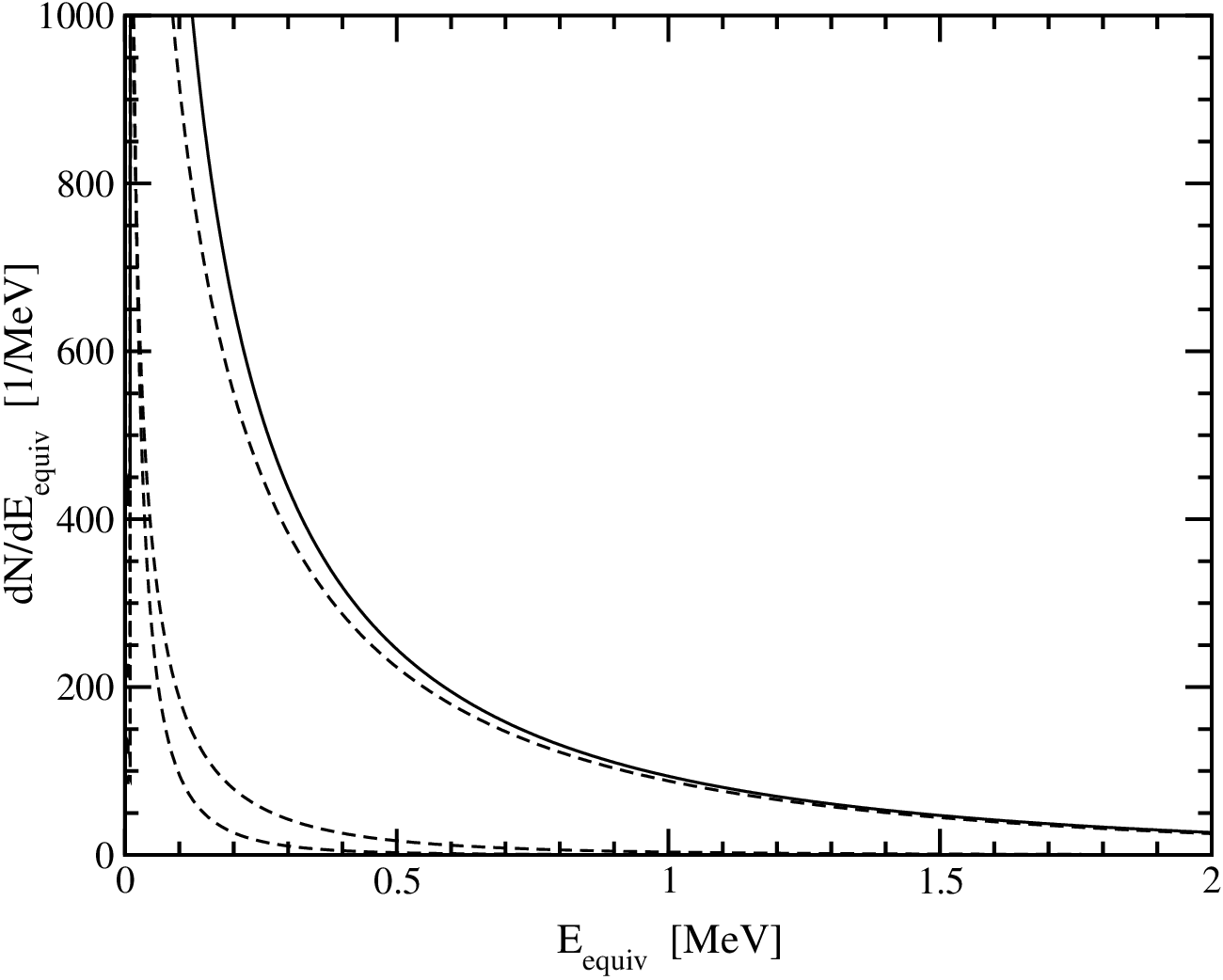}}\vspace{-1cm}
\caption{\label{fig:dndtquench} Analogous to Fig.~\ref{fig:dndt};
the struck proton spectrum for the different flavors, but with
quenching effects taken into account.  In order of increasing maximum
kinetic energy, the contributions from $\nu_e$, $\bar{\nu}_e$, and the
sum of $\nu_\mu$, $\nu_{\tau}$, $\bar{\nu}_{\mu}$, and
$\bar{\nu}_{\tau}$ are shown with dashed lines.  The solid line is the
sum spectrum for all flavors.  We assume a 1 kton detector mass for 
KamLAND.}
\end{figure} 
%%%%%%%%%%%%%%%%%%%%%%

Using the quenching function shown in Fig.~\ref{fig:quench}, we can
transform the true proton spectrum shown in Fig.~\ref{fig:dndt} into
the expected measured proton spectrum, shown in
Fig.~\ref{fig:dndtquench}.  If the quenching factor were a constant,
it would simply change the units of the $T_p$ axis.  However, it is
nonlinear, and reduces the light output of the lowest recoils the
most.  It also reduces the number of events above threshold.  The
anticipated threshold in KamLAND is 0.2 MeV electron equivalent
energy.  With the expected proton quenching, this corresponds to a
threshold on the true proton kinetic energy of 1.2 MeV.  The number of
events above this threshold for each flavor appears in
Table~\ref{tab:numevents}.  The measured proton spectrum will
primarily reflect the shape of the underlying Fermi-Dirac spectrum for
the sum of $\nu_{\mu}$, $\nu_{\tau}$ $\bar{\nu}_{\mu}$, and
$\bar{\nu}_{\tau}$.  This has been convolved with both the
differential cross section (which gives a range of $T_p$ for a given
$E_\nu$), and also the effects of quenching.  However, as we will
show, the properties of the initial neutrino spectrum can still be
reliably deduced.

%%%%%%%%%%%%%%%%%%%%%%
\begin{table}
\caption{Numbers of events in KamLAND (1 kton mass assumed) above the
noted thresholds for a standard supernova at 10 kpc, for the separate
flavors or their equivalents after oscillations.  Oscillations do not
change the number of neutrinos at a given energy, and the
neutral-current yields are insensitive to the neutrino flavor.
Equipartition among the six flavors is assumed (see the text for
discussion).  The thresholds are in electron equivalent energy, and
correspond to minimum true proton kinetic energies of 0 and 1.2 MeV.}
\begin{tabular}{l|rcr}
\it{Neutrino Spectrum} & $E_{thr} = 0$ & & $ 0.2$ MeV \\
\hline\hline
$\nu: T = 3.5$ MeV			& 57 	&& 3 	\\ \hline 
$\bar{\nu}: T = 5$ MeV			& 80 	&& 17 	\\ \hline 
$2 \nu: T = 8$ MeV			& 244	&& 127	\\ \hline 
$2 \bar{\nu}: T = 8$ MeV	 	& 243	&& 126	\\ \hline 
All 					& 624	&& 273	\\
\hline\hline
\end{tabular}
\label{tab:numevents}
\end{table}
%%%%%%%%%%%%%%%%%%%%%%

Background considerations, while important, are a small correction,
and so we ignore them here; see Ref.~\cite{elastic} for a complete
discussion.

%%%%%%%%%%%%%%%%%%%%%%%%%%%%%%%%%%%%%%%%%%%%%%%%%%%%%%%%%%%%%%%%%%%%%%%%%%%%%
%				Fits					    %
%%%%%%%%%%%%%%%%%%%%%%%%%%%%%%%%%%%%%%%%%%%%%%%%%%%%%%%%%%%%%%%%%%%%%%%%%%%%%

\section{Proton Spectrum Fits}
\label{fits}

The measured proton spectrum can be used to separately determine the
total energy of the $\nu_\mu$, $\nu_{\tau}$, $\bar{\nu}_{\mu}$, and
$\bar{\nu}_{\tau}$ neutrinos {\it and} their time-averaged
temperature.  The total number of detected events is proportional to
the portion of the total binding energy carried away by these four
flavors, and we denote this by $E^{tot}$ (note that this is {\it not}
the total binding energy $E_B$).  For a standard supernova, $E^{tot} =
4 (E_B/6) = 2/3 E_B \simeq 2 \times 10^{53}$ ergs.  We denote the
temperature of these four flavors by $T$.  If only the total yield
were measured, as for most neutral-current reactions, there would be
an unresolved degeneracy between $E^{tot}$ and $T$, since
\begin{equation}
N \sim E^{tot} \frac{\langle \sigma \rangle}{T}\,.
\end{equation}
Note that for $\sigma \sim E_\nu^n$, then $\langle \sigma \rangle \sim
T^n$.  For $\nu + d \rightarrow \nu + p + n$ in SNO, for example,
$\sigma \sim E^2$, so $N \sim E^{tot} T$.  Thus for a given
measured number of events, one would only be able to define a
hyperbola in the plane of $E^{tot}$ and $T$.  The scaling is
less simple here because of threshold effects, but the idea is the
same.

Here we have crucial information on the shape of the neutrino
spectrum, revealed through the proton spectrum.  To remind the reader,
in most neutral-current reactions there is {\it no} information on the
neutrino energy, e.g., one only counts the numbers of thermalized
neutron captures, or measures nuclear gamma rays (the energies of
which depend only on nuclear level splittings).

We performed quantitative tests of how well the parameters $E^{tot}$
and $T$ can be determined from the measured proton spectrum.  (We did
also investigate the effects of a chemical potential in the
Fermi-Dirac distribution, but found that it had little effect.  This
is simply because the cross section is not rising quickly enough to
see the tail of the thermal distribution in detail~\cite{SNsno}.)  Of
course, if the distance to the supernova is not known, then we are
effectively fitting for $E^{tot}/D^2$.

\begin{figure}
\centerline{\includegraphics[width=18pc]{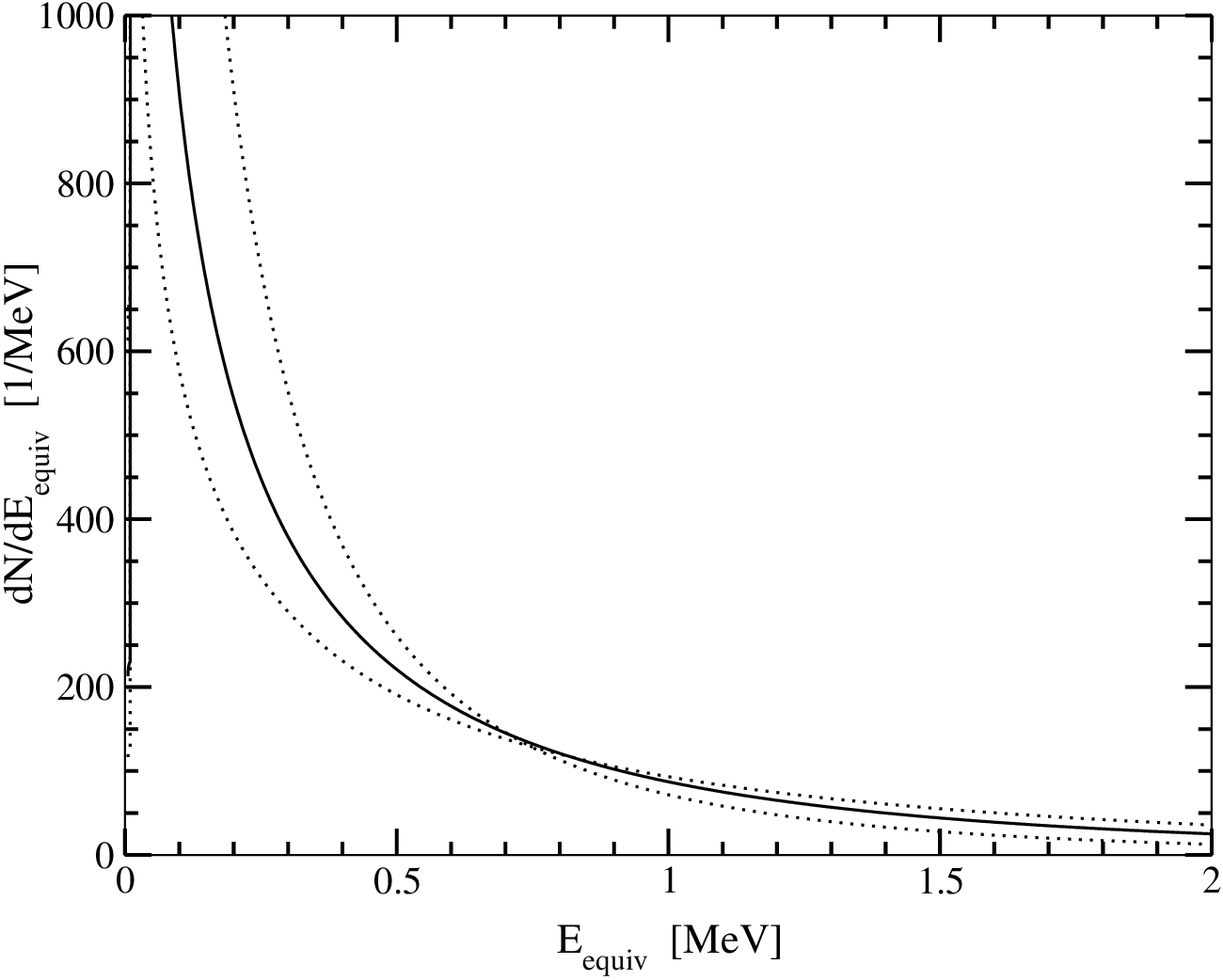}}\vspace{-1cm}
\caption{\label{fig:varspecs} Example spectra with different values of
$E^{tot}$ and $T$, all chosen to give the {\it same} number
of events above an electron equivalent threshold of 0.2 MeV (true
proton energy 1.2 MeV) in KamLAND.  Though not shown in this figure,
the spectrum above 2 MeV is included in our analysis.
At the 0.2 MeV point, from left to right
these correspond to $(E^{tot}, T)$ = (4.2, 6), (2.0, 8),
(1.4, 10), respectively, with $E^{tot}$ in $10^{53}$ ergs and
$T$ in MeV.}
\end{figure}

We performed Monte Carlo simulations of the supernova signal in
KamLAND and made chi-squared fits to determine $E^{tot}$ and
$T$ for each fake supernova.
To perform the fits, we started with an ``ideal'' spectrum, as
described by the integral:
\begin{eqnarray}
\label{idealspec}
\left(\frac{dN}{dT_p}\right)_{{\rm ideal}} & = &
C \int_0^\infty dT'_p\, G(T'_p; T_p, \delta T_p) \nonumber \\
& \times & \!\!\!\! \int_{(E_{\nu})_{min}}^\infty \!\! \hspace{-0.5cm}
dE_\nu\, f(E_\nu)\, \frac{d\sigma}{dT'_p}(E_\nu, T'_p)
\end{eqnarray}
where the inner integral is as in Eq.~(\ref{pspec}), with the addition
that quenching corrections are applied to $T'_p$ after convolution
with $f(E_\nu)$.  For the Gaussian energy resolution $G(T'_p; T_p,
\delta T_p)$, we used $\delta T_p = 0.1\sqrt{T_p/(1\ {\rm
MeV})}$~\cite{kamland}.  The normalization constant $C$ is given by
comparison to Eq.~(\ref{yield}).  Example spectra are shown in
Fig.~\ref{fig:varspecs}.

\begin{figure}
\centerline{\includegraphics[width=18pc]{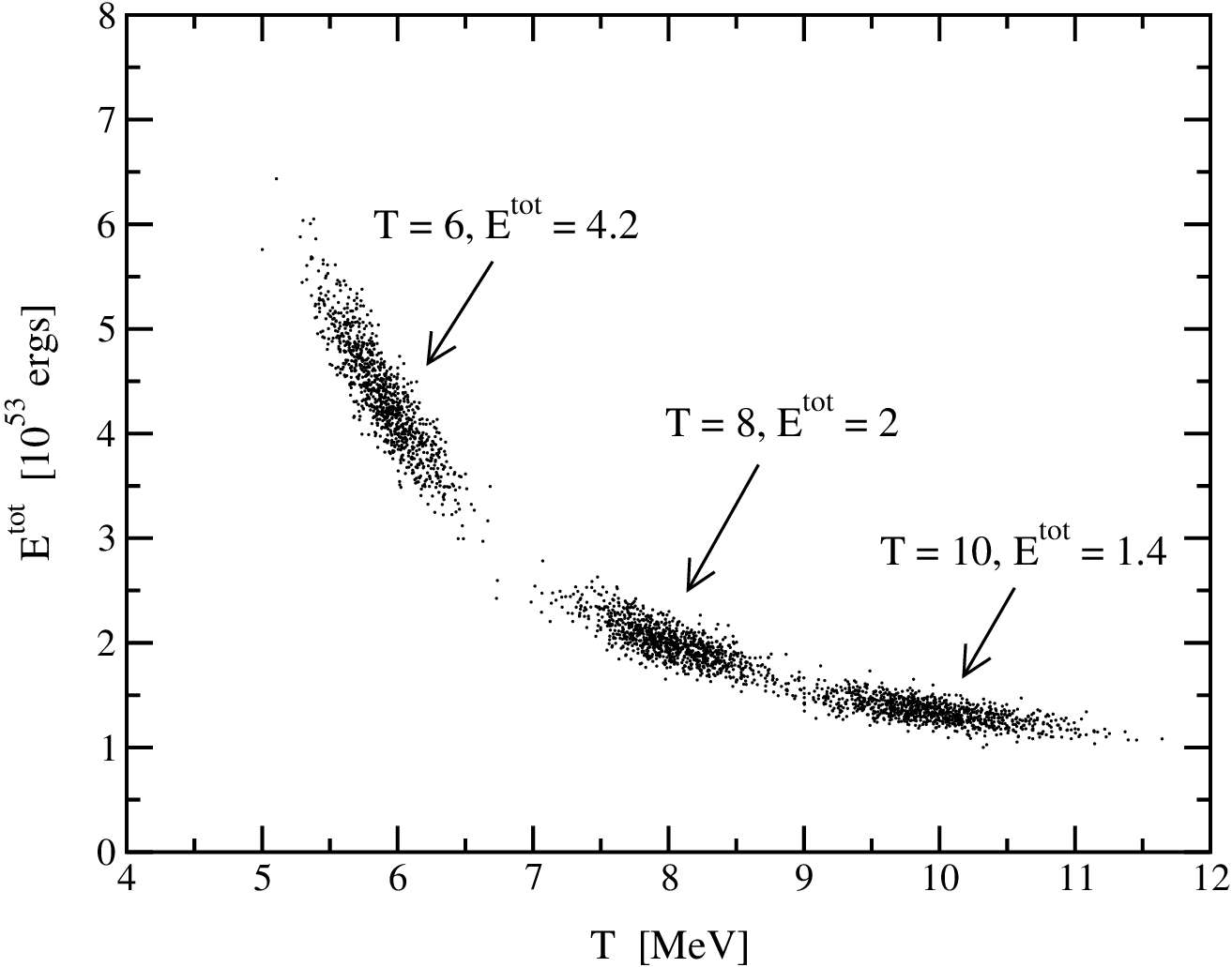}}\vspace{-1cm}
\caption{\label{fig:fits} Scatterplot of $10^3$ fitted values, in the
$E^{tot}$ and $T$ plane, for the labeled ``true'' values,
where $E^{tot}$ is the total portion of the binding energy
carried away by the sum of $\nu_\mu$, $\nu_{\tau}$, $\bar{\nu}_{\mu}$,
and $\bar{\nu}_{\tau}$, and $T$ is their temperature.  The values of
$E^{tot}$ and $T$ were chosen such that the numbers of
events above threshold were the same.  The measured shape of the
proton spectrum breaks the degeneracy between these two parameters.
Without that spectral information, one could not distinguish between
combinations of $E^{tot}$ and $T$ along the band in this
plane that our three example regions lie along.}
\end{figure}

Using $(dN/dT_p)_{\rm ideal}$, we binned the spectrum by the
following integral:
\begin{equation}
\label{binning}
N_i = \int_{(E_{min})_i}^{(E_{max})_i} dT_p
	\left(\frac{dN}{dT_p}\right)_{\rm ideal}
\end{equation}
where $N_i$ is the number of events in bin $i$, and $(E_{min})_i$ and
$(E_{max})_i$ are the minimum and maximum energies for bin $i$.
Eight bins of variable width were used, chosen to contain roughly
the same number of expected events per bin.
For a chosen $E^{tot}$ and $T$, this was the starting point
of our Monte Carlo (and the bin boundaries were kept fixed).
For each fake supernova, we sampled the number of
events in each of these bins according to the appropriate Poisson
distributions.  The resulting spectrum was as one might obtain from a
single supernova, given the finite number of events expected.  We then
varied $E^{tot}$ and $T$ in Eq.~(\ref{binning}) until the
values that best fit the fake spectral data were determined.  For a
given set of assumed $E^{tot}$ and $T$, this procedure was
repeated many times.  The distributions of the final
$E^{tot}$ and $T$ thus reveal the expected errors on fitting
$E^{tot}$ and $T$ for a single real future supernova.
Three examples are shown in Fig.~\ref{fig:fits}, where one can see
that $E^{tot}$ and $T$ can each be determined with
roughly 10\% error.

%%%%%%%%%%%%%%%%%%%%%%
%%%%%%%%%%%%%%%%%%%%%%

%%%%%%%%%%%%%%%%%%%%%%%%%%%%%%%%%%%%%%%%%%%%%%%%%%%%%%%%%%%%%%%%%%%%%%%%%%%%%
%			Conclusions					    %
%%%%%%%%%%%%%%%%%%%%%%%%%%%%%%%%%%%%%%%%%%%%%%%%%%%%%%%%%%%%%%%%%%%%%%%%%%%%%

\section{Discussion and Conclusions}

We have shown that neutrino-proton elastic scattering, previously
unrecognized as a useful detection reaction for low-energy neutrinos,
in fact has a yield for a supernova comparable to $\bar{\nu}_e + p
\rightarrow e^+ + n$, even after taking into account the quenching of
the proton scintillation light and assuming a realistic detector
threshold.

In addition, the measured proton spectrum is related to
the incident neutrino spectrum.  We have shown explicitly that one can
separately measure the total energy and temperature of
$\nu_\mu$, $\nu_\tau$, $\bar{\nu}_\mu$, and $\bar{\nu}_\tau$, each
with uncertainty of order 10\% in KamLAND.  This greatly enhances the 
importance of detectors like KamLAND and Borexino for detecting supernova
neutrinos.

%%%%%%%%%%%%%%%%%%%%
%%%%%%%%%%%%%%%%%%%%%

For Borexino, the useful volume for supernova neutrinos is 0.3 kton,
and the hydrogen to carbon ratio in the pure pseudocumene
(C$_9$H$_{12}$) is $1.3:1$~\cite{SNborexino}, so there are about 4.7
times fewer free proton targets than assumed for KamLAND.  However,
the quenching is less in pure scintillator (KamLAND is about 20\%
pseudocumene and 80\% paraffin oil~\cite{kamland}), and the errors on
$E^{tot}$ and $T$ scale as $1/\sqrt{N}$, so that the precision in
Borexino should be about 20\% or better.

Other techniques for bolometric measurements of supernova neutrino
fluxes have been studied.  Detectors for elastic neutral-current
neutrino scattering on electrons~\cite{cabrera} and coherently on
whole nuclei~\cite{coherent} have been discussed, but never built.
If neutrino oscillations are effective in swapping spectra, then the
temperature of the ``hot'' flavors may be revealed in the measured
positron spectrum from $\bar{\nu}_e + p \rightarrow e^+ + n$; two
recent studies have shown very good precision ($< 5\%$) for
measuring the temperatures and the total binding
energy~\cite{barger,minakata}.  However, they assumed exact energy
equipartition among the six neutrino flavors, whereas the uncertainty
on equipartition is at least 50\%~\cite{raffeltproc}.  Nevertheless,
under less restrictive assumptions, this technique may play a
complementary role.  Finally, since for different cross sections, the
neutral-current yields depend differently on temperature, comparison
of the yields may provide some information~\cite{relyields}.  However,
there are caveats.  In neutrino-electron scattering, the neutrino
energy is not measured because the neutrino-electron angle is much
less than the angular resolution due to multiple scattering.  The
scattered electrons, even those in a forward cone, sit on a much
larger background of $\bar{\nu}_e + p \rightarrow e^+ + n$ events, so
it is difficult to measure their spectrum~\cite{SNpoint}; also, their
total yield is only weakly dependent on temperature.  At the other
extreme (see Fig.~3 of Ref.~\cite{relyields}), the yield of
neutral-current events~\cite{SNsk} on $^{16}$O depends strongly on a
possible chemical potential term in the thermal distribution.

It is important to note that the detection of recoil protons from {\it
neutron}-proton elastic scattering at several MeV has been routinely
accomplished in scintillator detectors (see, e.g.,
Ref.~\cite{routine}).  Since both particles are massive, the proton
will typically take half of the neutron energy.  This reaction
provides protons in the same energy range as those struck in
neutrino-proton elastic scattering with $E_\nu \sim 30$ MeV.  This is
a very important proof of concept for all aspects of the detection of
low-energy protons.

Though low-energy backgrounds will be challenging, it is also
important to note that the background requirements for detecting the
supernova signal are approximately 3 orders of magnitude {\it less}
stringent than those required for detecting solar neutrinos in the
same energy range (taking quenching into account for our signal).
Borexino has been designed to detect very low-energy solar neutrinos,
and KamLAND hopes to do so in a later phase of the experiment.

These measurements would be considered in combination with similar
measurements for $\nu_e$ and $\bar{\nu}_e$ from charged-current
reactions in other detectors.  Separate measurements of the total
energy and temperature for each flavor will be invaluable for
comparing to numerical supernova models~\cite{SNmodels,SNnew}.  They
will also be required to make model-independent studies of the effects
of neutrino oscillations~\cite{SNosc}.  If the total energy release
$E_B$ in all flavors has been measured, then
\begin{equation}
E_B \simeq \frac{3}{5} \frac{G M_{NS}^2}{R_{NS}}\,,
\end{equation}
thus allowing a direct and unique measurement of the newly-formed 
neutron star properties, principally the mass $M_{NS}$~\cite{NS}.

%%%%%%%%%%%%%%%%%%%%%%%%%%%%%%%%%%%%%%%%%%%%%%%%%%%%%%%%%%%%%%%%%%%%%%%
%			Acknowledgements			      %
%%%%%%%%%%%%%%%%%%%%%%%%%%%%%%%%%%%%%%%%%%%%%%%%%%%%%%%%%%%%%%%%%%%%%%%

\section*{ACKNOWLEDGEMENTS}
This talk was based on collaborative work with Will Farr and Petr
Vogel, whom I happily acknowledge.  JFB is supported as the David
N. Schramm Fellow at Fermilab, which is operated by URA under DOE
contract No. DE-AC02-76CH03000.  JFB was additionally supported by
NASA under NAG5-10842.

%%%%%%%%%%%%%%%%%%%%%%%%%%%%%%%%%%%%%%%%%%%%%%%%%%%%%%%%%%%%%%%%%%%%%%%%%%%%%%
%			Bibliography					     %
%%%%%%%%%%%%%%%%%%%%%%%%%%%%%%%%%%%%%%%%%%%%%%%%%%%%%%%%%%%%%%%%%%%%%%%%%%%%%%


\begin{thebibliography}{100}

\bibitem{elastic}
J.~F.~Beacom, W.~M.~Farr and P.~Vogel,
Phys.\ Rev.\ D {\bf 66}, 033001 (2002).
%%CITATION = HEP-PH 0205220;%%

\bibitem{weinberg}
S.~Weinberg, Phys.\ Rev.\ D {\bf 5}, 1412 (1972).
%%CITATION = PHRVA,D5,1412;%%

\bibitem{ahrens}
L.~A.~Ahrens {\it et al.}, Phys.\ Rev.\ D {\bf 35}, 785 (1987).
%%CITATION = PHRVA,D35,785;%%

\bibitem{sigma}
C.~H.~Llewellyn Smith, Phys.\ Rept.\  {\bf 3}, 261 (1972);
%%CITATION = PRPLC,3,261;%%
S.~M.~Bilenky and J.~Hosek, Phys.\ Rept.\  {\bf 90}, 73 (1982).
%%CITATION = PRPLC,90,73;%%

\bibitem{SNmodels}
J.R.~Wilson and R.W.~Mayle, Phys.\ Rept.\ {\bf 227}, 97 (1993);
%%CITATION = PRPLC,227,97;%%
M. Herant, W. Benz, W.R. Hix, C.L Fryer, and S.A. Colgate, 
Astrophys. J. {\bf 435}, 339 (1994);
%%CITATION = ASJOA,435,339;%%
A. Burrows, J. Hayes, and B.A. Fryxell, 
Astrophys.\ J.\  {\bf 450}, 830 (1995);
%%CITATION = ASJOA,450,830;%%
M.~Rampp and H.~T.~Janka, Astrophys.\ J.\  {\bf 539}, L33 (2000);
%%CITATION = ASTRO-PH 0005438;%%
A.~Mezzacappa, M.~Liebendorfer, O.~E.~Messer, W.~R.~Hix,
F.~K.~Thielemann and S.~W.~Bruenn, 
Phys.\ Rev.\ Lett.\  {\bf 86}, 1935 (2001);
%%CITATION = ASTRO-PH 0005366;%%
C.L. Fryer, A. Heger, Astrophys.\ J.\ {\bf 541}, 1033 (2000);
%%CITATION = ASTRO-PH 9907433;%%
C.~J.~Horowitz, Phys.\ Rev.\ D {\bf 65}, 043001 (2002);
%%CITATION = ASTRO-PH 0109209;%%
M.~Rampp and H.~T.~Janka, arXiv:astro-ph/0203101;
%%CITATION = ASTRO-PH 0203101;%%
M.~Liebendoerfer, O.~E.~Messer, A.~Mezzacappa, S.~W.~Bruenn,
C.~Y.~Cardall and F.~K.~Thielemann, arXiv:astro-ph/0207036.
%%CITATION = ASTRO-PH 0207036;%%

\bibitem{raffeltproc}
G.~G.~Raffelt, hep-ph/0201099.
%%CITATION = HEP-PH 0201099;%%

\bibitem{SNrate}
J.~F.~Beacom, R.~N.~Boyd and A.~Mezzacappa,
Phys.\ Rev.\ D {\bf 63}, 073011 (2001).
%%CITATION = ASTRO-PH 0010398;%%

\bibitem{SNmb}
M.~K.~Sharp, J.~F.~Beacom and J.~A.~Formaggio,
Phys.\ Rev.\ D {\bf 66}, 013012 (2002).
%%CITATION = HEP-PH 0205035;%%

\bibitem{invbeta}
P.~Vogel and J.~F.~Beacom, Phys.\ Rev.\ D {\bf 60}, 053003 (1999).
%%CITATION = HEP-PH 9903554;%%

\bibitem{kamland} 
A. Suzuki, in {\it Lepton and Baryon Number Violation Particle
Physics, Astrophysics, and Cosmology}, eds. H.V. Klapdor-Kleingrothaus
and I.V. Krivosheina (Institute of Physics, Philadelphia, 1999);
S.~J. Freedman, G.~Gratta, {\it et al.}, {\it Proposal for US
Participation in KamLAND}, March 1999;
A.~Piepke, Nucl.\ Phys.\ Proc.\ Suppl.\  {\bf 91}, 99 (2001).
%%CITATION = NUPHZ,91,99;%%

\bibitem{SNsno}
J.~F.~Beacom and P.~Vogel, Phys.\ Rev.\ D {\bf 58}, 093012 (1998).
%%CITATION = HEP-PH 9806311;%%

\bibitem{SNborexino}
L.~Cadonati, F.~P.~Calaprice and M.~C.~Chen,
Astropart.\ Phys.\  {\bf 16}, 361 (2002).
%%CITATION = HEP-PH 0012082;%%

\bibitem{cabrera}
B.~Cabrera, L.~M.~Krauss and F.~Wilczek,
Phys.\ Rev.\ Lett.\  {\bf 55}, 25 (1985).
%%CITATION = PRLTA,55,25;%%

\bibitem{coherent}
D.~Z.~Freedman, Phys.\ Rev.\ D {\bf 9}, 1389 (1974);
%%CITATION = PHRVA,D9,1389;%%
A.~Drukier and L.~Stodolsky, Phys.\ Rev.\ D {\bf 30}, 2295 (1984).
%%CITATION = PHRVA,D30,2295;%%

\bibitem{barger}
V.~Barger, D.~Marfatia and B.~P.~Wood, arXiv:hep-ph/0112125.
%%CITATION = HEP-PH 0112125;%%

\bibitem{minakata}
H.~Minakata, H.~Nunokawa, R.~Tomas and J.~W.~Valle,
Phys.\ Lett.\ B {\bf 542}, 239 (2002).
%%CITATION = HEP-PH 0112160;%%

\bibitem{relyields} 
J.~F. Beacom, in {\it Neutrinos in the New Millennium}, eds.
G. Domokos and S. Kovesi-Domokos (World Scientific, Singapore, 2000),
hep-ph/9909231.  Note that the version in the proceedings volume 
contains no minus signs; the version on the arXiv is correct.
%%CITATION = HEP-PH 9909231;%%

\bibitem{SNpoint}
J.~F.~Beacom and P.~Vogel, Phys.\ Rev.\ D {\bf 60}, 033007 (1999).
%%CITATION = ASTRO-PH 9811350;%%

\bibitem{SNsk}
K.~Langanke, P.~Vogel and E.~Kolbe, 
Phys.\ Rev.\ Lett.\  {\bf 76}, 2629 (1996);
%%CITATION = NUCL-TH 9511032;%%
J.~F.~Beacom and P.~Vogel, Phys.\ Rev.\ D {\bf 58}, 053010 (1998).
%%CITATION = HEP-PH 9802424;%%

\bibitem{routine}
J.~B. Czirr, D.~R. Nygren, and C.~D. Zafiratos,
Nucl.\ Instrum.\ Meth.\  {\bf 31}, 226 (1964);
%%CITATION = NUIMA,31,226;%%
K.~H. Maier and J. Nitschke,
Nucl.\ Instrum.\ Meth.\  {\bf 59}, 227 (1968);
%%CITATION = NUIMA,59,227;%%
R.~L. Craun and D.~L. Smith,
Nucl.\ Instrum.\ Meth.\  {\bf 80}, 239 (1970);
%%CITATION = NUIMA,80,239;%%
D.~J.~Ficenec, S.~P.~Ahlen, A.~A.~Marin, J.~A.~Musser and G.~Tarle,
Phys.\ Rev.\ D {\bf 36}, 311 (1987);
%%CITATION = PHRVA,D36,311;%%
S.~Mouatassim, G.~J.~Costa, G.~Guillaume, B.~Heusch, A.~Huck and M.~Moszynski,
Nucl.\ Instrum.\ Meth.\  {\bf 359A}, 530 (1995);
%%CITATION = NUIMA,359A,530;%%
B.~Achkar {\it et al.}, Phys.\ Lett.\ B {\bf 374}, 243 (1996).
%%CITATION = PHLTA,B374,243;%%

\bibitem{SNnew}
M.~T.~Keil, G.~G.~Raffelt and H.~T.~Janka, arXiv:astro-ph/0208035;
%%CITATION = ASTRO-PH 0208035;%%
G.~G.~Raffelt, arXiv:astro-ph/0105250;
%%CITATION = ASTRO-PH 0105250;%%
R.~Buras, H.~T.~Janka, M.~T.~Keil, G.~G.~Raffelt and M.~Rampp,
arXiv:astro-ph/0205006;
%%CITATION = ASTRO-PH 0205006;%%

\bibitem{SNosc}
Refs.~\cite{barger,minakata},
B.~Jegerlehner, F.~Neubig and G.~Raffelt,
Phys.\ Rev.\ D {\bf 54}, 1194 (1996);
%%CITATION = ASTRO-PH 9601111;%%
E.~K.~Akhmedov, C.~Lunardini and A.~Y.~Smirnov, arXiv:hep-ph/0204091;
%%CITATION = HEP-PH 0204091;%%
C.~Lunardini and A.~Y.~Smirnov, Nucl.\ Phys.\ B {\bf 616}, 307 (2001);
%%CITATION = HEP-PH 0106149;%%
A.~S.~Dighe and A.~Y.~Smirnov, Phys.\ Rev.\ D {\bf 62}, 033007 (2000);
%%CITATION = HEP-PH 9907423;%%
R.~C.~Schirato, G.~M.~Fuller, arXiv:astro-ph/0205390;
%%CITATION = ASTRO-PH 0205390;%%
G.~L.~Fogli, E.~Lisi, D.~Montanino and A.~Palazzo,
Phys.\ Rev.\ D {\bf 65}, 073008 (2002)
[Erratum-ibid.\ D {\bf 66}, 039901 (2002)];
%%CITATION = HEP-PH 0111199;%%
K.~Takahashi, M.~Watanabe, K.~Sato and T.~Totani,
Phys.\ Rev.\ D {\bf 64}, 093004 (2001);
%%CITATION = HEP-PH 0105204;%%
G.~C.~McLaughlin, J.~M.~Fetter, A.~B.~Balantekin and G.~M.~Fuller,
Phys.\ Rev.\ C {\bf 59}, 2873 (1999);
%%CITATION = ASTRO-PH 9902106;%%
J.~Fetter, G.~C.~McLaughlin, A.~B.~Balantekin and G.~M.~Fuller,
arXiv:hep-ph/0205029;
%%CITATION = HEP-PH 0205029;%%
M.~Sorel and J.~Conrad, Phys.\ Rev.\ D {\bf 66}, 033009 (2002);
%%CITATION = HEP-PH 0112214;%%
G.~Dutta, D.~Indumathi, M.~V.~Murthy and G.~Rajasekaran,
Phys.\ Rev.\ D {\bf 61}, 013009 (2000);
%%CITATION = HEP-PH 9907372;%%
H.~Ejiri, J.~Engel and N.~Kudomi, Phys.\ Lett.\ B {\bf 530}, 27 (2002);
%%CITATION = ASTRO-PH 0112379;%%
S.~Pastor and G.~Raffelt, arXiv:astro-ph/0207281.
%%CITATION = ASTRO-PH 0207281;%%

\bibitem{NS}
J.~M.~Lattimer and M.~Prakash, Astrophys.\ J.\  {\bf 550}, 426 (2001).
%%CITATION = ASTRO-PH 0002232;%%

\end{thebibliography}
\end{document}